\documentclass[conference,a4paper]{IEEEtran}

\usepackage{amsmath,amssymb,graphicx}
\usepackage{cite}
\usepackage{hyperref}
\usepackage{tikz}
\usetikzlibrary{graphs,graphs.standard,quotes,calc}

\title{The Coherent Multiplex: Scalable Real-Time Wavelet Coherence Architecture}
\author{\IEEEauthorblockN{Noah Shore}
\IEEEauthorblockA{\textit{School Of Mathematical and Statistical Sciences} \\
\textit{University of Galway}\\
Galway, Ireland \\
n.shore2@universityofgalway.ie}}
\def\BibTeX{{\rm B\kern-.05em{\sc i\kern-.025em b}\kern-.08em
    T\kern-.1667em\lower.7ex\hbox{E}\kern-.125emX}}
\begin{document}
\maketitle
\date{August 2025}

\begin{abstract}

The Coherent Multiplex is formalized and validated as a scalable, real-time system for identifying, analyzing, and visualizing coherence among multiple time series. Its architecture comprises a fast spectral similarity layer based on cosine similarity metrics of Fourier-transformed signals, and a sparse time-frequency layer for wavelet coherence. The system constructs and evolves a multilayer graph representing inter-signal relationships, enabling low-latency inference and monitoring. A simulation prototype demonstrates functionality across 8 synthetic channels with a high similarity threshold for further computation, with additional opportunities for scaling the architecture up to support thousands of input signals with constrained hardware. Applications discussed include neuroscience, finance, and biomedical signal analysis.
\end{abstract}


\section{Introduction}

Understanding and employing wavelet coherence methods between two signals has been increasing in popularity in fields like neuroscience, finance, geophysics, and communications, where uncovering time-dependent relationships between non-stationary signals can reveal underlying mechanisms and obscured correlations. Conventional approaches to coherence analysis are limited to high-level programming languages, run outdated transforms, and are incapable of handling high-volumes or real-time data streams efficiently. The Coherent Multiplex overcomes these challenges by offering a rapid and scalable solution for live coherence tracking, allowing users to explore frequency- and time-dependent interactions with greater speed and precision.

By leveraging recent algorithms and optimized computing frameworks in the \textit{Fastest Fourier Transform in the West} (FFTW)~\cite{frigo} and the \textit{Fast Continuous Wavelet Transform} (FCWT)~\cite{arts}, the Multiplex significantly reduces the processing time required for coherence analysis, making it suitable for time-dependent and large-scale applications. This innovation, combined with the living multilayered network architecture, enables researchers and industry professionals to efficiently monitor and investigate dynamic signal relationships as they evolve, supporting new opportunities in data-driven decision making and responsive system design across diverse scientific and industrial domains.

\section{Related Work}

Wavelet coherence has been applied in neuroscience \cite{QASSIM201323}, econometrics \cite{VUKOVIC2021101457}, and climatology \cite{grinsted2004application} to study dynamic relationships between nonstationary signals. Traditional implementations rely on post-hoc analysis of recorded signals, using tools contained in MATLAB, R, or Python environments. These packages emphasize flexibility but are not designed for streaming or high-throughput applications, and their computational costs make them impractical for monitoring at scale.

Real-time coherence monitoring is less common. Systems such as \cite{mi_brain_msc} support streaming \textit{magnitude-squared coherence} (MSC) in specialized domains such as brain–computer interfaces (BCIs), but this is a much more superficial metric than wavelet coherence. Other streaming signal analysis tools, like LabStreamingLayer \cite{kothe}, focus on synchronization and data collection rather than coherence analysis.

Network representations of signal similarity are also established in neuroscience and finance, where pairwise correlations or coherence values are used to construct functional connectivity or market dependency graphs. However, these approaches are typically retrospective, relying on fully collected datasets. To our knowledge, no existing technology offers an integrated pipeline for continuous ingestion, transformation, and multi-layered network construction of FFT and wavelet coherence with the extensibility and modularity of the Multiplex.

\section{System Overview}

\begin{enumerate}

\item

The pipeline supports flexible signal ingestion through APIs or direct connections to monitoring devices. This allows users to seamlessly integrate diverse data sources, whether signals are generated internally or are continuous real-world data streams.

\item

To efficiently analyze signals in the frequency domain, Coherent Multiplex employs fast Fourier transforms computed simultaneously using optimized libraries. This approach ensures rapid processing of high-throughput data streams, facilitating immediate insight into the spectral characteristics of each signal.

\item

A multilayered graph network is constructed to represent the complex relationships among signals. Each node corresponds to an individual signal, and edges denote the strength of coherence or similarity between them. This layered structure supports exploration at multiple levels of granularity and aids in visualizing interdependencies.

\item

Cosine similarity metrics are calculated pairwise between FFT magnitude vectors in the frequency domain. This measurement quantifies the alignment of spectral content between signals, serving as a computationally efficient heuristic to identify pairs exhibiting potential coherence, which can then be further analyzed.

\item

Additional edges between nodes will grow and decay based on cosine similarity. These edges contain wavelet coherence arrays that are updated at a subsampled interval because of prohibitive cost. Utilizing continuous wavelet transforms, this method provides detailed time-frequency resolution, revealing dynamic and transient coherent interactions between signals that evolve over time.

\item

An interface dashboard presents a visualization suite including time-domain signal plots, FFT spectra, network graphs illustrating signal relationships, and wavelet coherence heatmaps. These elements can be monitored with live updates delivered via server-sent events.
\end{enumerate}

\section{Methodology}

\subsection{Signal Buffering}

We consider \(M\) discrete-time signals sampled at frequency \(f_s\), each represented by a sliding buffer of length \(N\):
\[
X_i = \bigg[x_i[n-N+1], \dotsc, x_i[n]\bigg], \quad i = 1, \dotsc, M.
\]
Each buffer contains the most recent \(N\) samples up to and including the current time \(n\), ensuring causality for real-time processing.

These buffers are arranged into a data matrix \(D \in \mathbb{R}^{M \times N} \), where
\[
D =
\begin{bmatrix}
x_1[n-N+1] & x_1[n-N+2] & \dots & x_1[n] \\
x_2[n-N+1] & x_2[n-N+2] & \dots & x_2[n] \\
\vdots & \vdots & \ddots & \vdots \\
x_M[n-N+1] & x_M[n-N+2] & \dots & x_M[n]
\end{bmatrix}.
\]
This matrix serves as the input for subsequent frequency-domain analysis.

\subsection{Spectral Cosine Similarity}

We apply the Discrete Fourier Transform (DFT) to each real-valued signal buffer to obtain its frequency-domain representation. For each \(X_i \in \mathbb{R}^N\), define:
\[
F_i[k] = \sum_{n=0}^{N-1} X_i[n] e^{-2\pi \mathbf{j} k n / N}, \quad k = 0, \dotsc, N-1.
\]
Where \( \mathbf{j} = \sqrt{-1} \). Since each \(X_i\) is real, its DFT exhibits Hermitian symmetry: \(F_i[N - k] = \overline{F_i[k]}\). As a result, the spectral content is fully captured in the first \(L = \frac{N}{2}+ 1\) frequencies (assuming \(N\) even), and we discard redundant components.

The truncated frequency vectors are stacked into a matrix \(F \in \mathbb{C}^{M \times L}\), where each row corresponds to the spectrum of one signal.

Parseval’s identity ensures energy is preserved across domains, so we compute cosine distance directly in the frequency domain:
\[
d(F_i, F_j) = \frac{|\langle F_i, F_j \rangle|}{\|F_i\| \cdot \|F_j\|}.
\]
This yields a symmetric similarity matrix \(S \in \mathbb{R}^{M \times M}\).

Here, \(\langle F_i, F_j \rangle = \sum_{k=1}^N F_i[k] \, \overline{F_j[k]}\) denotes the standard inner product of the frequency-domain vectors, and \(\|F_i\| = \sqrt{\langle F_i, F_i \rangle}\) is the corresponding Euclidean (\(\ell_2\)) norm.

\subsection{Exploratory Network}
The pairwise cosine similarity matrix \( S \in \mathbb{R}^{M \times M} \) defines a weighted, undirected graph \( G_1 = (V, E_1, W_1) \), where each vertex \( v_i \in V \) corresponds to a signal \( i \), and each edge \( e_{ij} \in E_1, i \not=j \) represents the spectral similarity between signals \( i \) and \( j \). Specifically, edges are assigned weights based on the similarity function
\[
W_1(i,j) = d(i,j), \quad d(i,j) \in [0,1],
\]
where \( d(i,j) \) reflects the cosine similarity between the frequency-domain representations of signals \( i \) and \( j \). A value of \( d(i,j) = 1 \) indicates maximal similarity (i.e., parallel vectors in the spectral space), while \( d(i,j) = 0 \) indicates orthogonality.

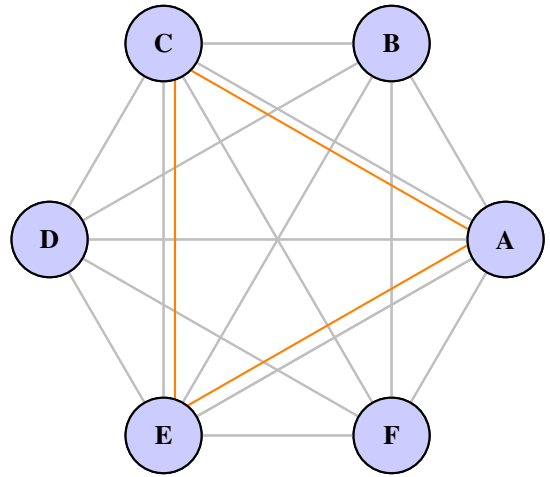
\begin{figure}[b]
\centering
\begin{tikzpicture}[main node/.style={circle,draw,thick,fill=blue!20,minimum size=1cm, font=\bfseries}]

\foreach \angle/\name in {0/A, 60/B, 120/C, 180/D, 240/E, 300/F}
  \node[main node] (\name) at ({3*cos(\angle)}, {3*sin(\angle)}) {\name};

\foreach \x in {A,B,C,D,E,F} {
  \foreach \y in {A,B,C,D,E,F} {
    \ifx\x\y
    \else
      \draw[gray!50, thick] (\x) -- (\y);
    \fi
  }
}

\draw[gray!50, thick] (A) -- (E);

\path let \p1 = (A), \p2 = (E) in 
    coordinate (shiftA) at (\x1,\y1) 
    coordinate (shiftB) at (\x2,\y2);

\draw[orange, thick] 
    ($(shiftA) + (0.15,0.3)$) -- ($(shiftB) + (0.15,0.3)$);

\path let \p1 = (A), \p2 = (C) in
    coordinate (shiftA) at (\x1,\y1)
    coordinate (shiftC) at (\x2,\y2);
\draw[orange, thick] 
    ($(shiftA) + (0.35,-0.35)$) -- ($(shiftC) + (0.35,-0.35)$);

\path let \p1 = (C), \p2 = (E) in
    coordinate (shiftC) at (\x1,\y1)
    coordinate (shiftE) at (\x2,\y2);
\draw[orange, thick] 
    ($(shiftC) + (0.15,-0.15)$) -- ($(shiftE) + (0.15,-0.15)$);

\foreach \angle/\name in {0/A, 60/B, 120/C, 180/D, 240/E, 300/F}
  \node[main node] (\name) at ({3*cos(\angle)}, {3*sin(\angle)}) {\name};
  
\end{tikzpicture}
\caption{Combined network showing two layers for a six-signal dataset. Fully connected layer $G_1$ (gray edges) represents pairwise spectral similarity, while sparse layer $G_2$ (orange edges) contains coherence arrays.}
\label{fig:combinedgraph}
\end{figure}

By construction, \( G_1 \) is a fully connected (complete) graph, capturing all pairwise relationships among the \( M \) signals. This graph provides a compact and interpretable structure for understanding global spectral relationships within the dataset. The weight matrix \( W_1 \) can be interpreted as a similarity kernel, suitable for downstream tasks such as clustering, dimensionality reduction, or community detection. Fig.~\ref{fig:combinedgraph} displays an example where \(M=6\).

\subsection{Depth Layer}

To capture time-localized, frequency-specific relationships between signals, we construct a second, sparse network layer \( G_2 = (V, E_2, W_2) \) where edges represent \textit{wavelet coherence}.

Given a time series \( x \), its wavelet transform~\cite{torrence1998practical} is defined as
\[
W_x(t, s) = \sum_{n=0}^{N-1} x[n] \cdot \overline{\psi} \left( \frac{n - t}{s} \right),
\]
where \( \psi \) is a complex-valued mother wavelet, \( s \) is the scale parameter (related to frequency), and \( t \) is the time index.

The wavelet coherence~\cite{grinsted2004application} between signals \( x_i \) and \( x_j \) at time \( t \) and scale \( s \) is
\[
C_{ij}(t, s) = \frac{\left| S\left( W_i(t, s) \cdot \overline{W_j(t, s)} \right) \right|^2}
{S\left( |W_i(t, s)|^2 \right) \cdot S\left( |W_j(t, s)|^2 \right)},
\]
where \( S(\cdot) \) is a smoothing operator in time and scale, and \( C_{ij}(t, s) \in [0,1] \).

Edges in \( G_2 \) exist only between pairs of signals with coherence exceeding a threshold \( \theta \in [0,1] \), so
\[
e_{ij} \in E_2 \iff W_1(i,j) \geq \theta,
\]
and the corresponding edge weight is
\[
W_2(i,j) = C_{ij}(t, s).
\]

This sparsification ensures \( G_2 \) highlights significant localized interactions, modulated by the spectral similarities encoded in \( G_1 \). For example, if edges \((A,C)\), \((C,E)\), and \((A,E)\) have weights above \( \theta \), the resulting layer \( G_2 \) appears alongside \(G_1\) in Fig.~\ref{fig:combinedgraph}.

\section{Computation \& Performance}

We denote the number of signals by $M$, the buffer length (samples) by $N$, the number of retained FFT bins by $L = \lfloor N/2 \rfloor + 1$, and the number of wavelet scales by $Q$.

\subsection{Pipeline Steps and Costs (Per Update)}

\begin{enumerate}
    \item \textbf{Signal buffering}:
    Append the newest sample into circular buffers of length $N$.
    Time complexity is $O(1)$ per signal (circular buffer) or $O(M)$ if all buffers are touched.
    Memory: $O(MN)$ samples.
    
    \item \textbf{FFT}:
    Per signal: $O(N\log N)$ using an FFT library (FFTW).
    All signals: $O(M\,N\log N)$.
    With real input, only $L$ bins are needed downstream.
    
    \item \textbf{Magnitude / feature extraction}:
    Extract magnitudes from each FFT output: $O(M L)$.
    
    \item \textbf{Pairwise cosine similarities}:
    Naïve: $\binom{M}{2}$ dot products of length $L$, giving $O(M^2 L)$.
    Norms: $O(M L)$.
    Similarity matrix storage: $O(M^2)$.
    
    \item \textbf{Candidate selection / sparsification}:
    Threshold $S$ to produce $K$ candidate edges with $W_1(i,j) \ge \theta$: $O(M^2)$.
    
    \item \textbf{Continuous wavelet transform}:
    FFT-based CWT at $Q$ scales: $O(Q\,N\log N)$ per signal.
    If only $P$ signals are active candidate pairs, the cost is $O(P\,Q\,N\log N)$.
    
    \item \textbf{Wavelet coherence}:
    Given $W_i, W_j \in \mathbb{C}^{Q \times N}$, coherence computation is $ O(QN)$.
\end{enumerate}

While FFTs and cosine similarity computations scale to thousands of signals in real time, wavelet coherence is computed selectively on candidate pairs filtered by the similarity threshold. This design allows the architecture to scale up to large corpora while limiting the computation expenditure at the coherence.

\subsection{Benchmarking}

Memory consumption for FFT and cosine similarity metrics is well-documented at scale, and the multiplex does not rely on novel implementations for these steps. Our contribution lies in a novel implementation of the coherence layer. To evaluate its performance, we measure the compute time required to generate coherence arrays for signal pairings across varying sample sizes and scale resolutions.

We compare our implementation against the PyCWT library's wavelet coherence transform (WCT) \cite{krieger} using two synthetic noise signals sampled at 8000 Hz. Figure~\ref{perf} shows the impact of increasing frequency and time resolution on memory usage. The results closely mirror the efficiency gains reported in the original FCWT publication~\cite{arts}, which is the wavelet transform engine behind our implementation. Note the step-wise behavior observed with increasing signal duration is characteristic of FFT-based methods, which are most efficient for input lengths of $2^n, n \in \mathbb{N}$. Consequently, such algorithms typically pad samples to the next power of 2 \cite{smith}.

\definecolor{mplblue}{rgb}{0.121, 0.466, 0.705}
\definecolor{mplorange}{rgb}{1.0, 0.498, 0.054}

\begin{figure}[t]
    \centering
    \includegraphics[trim=0 0 0 0,clip,width=.49\textwidth]{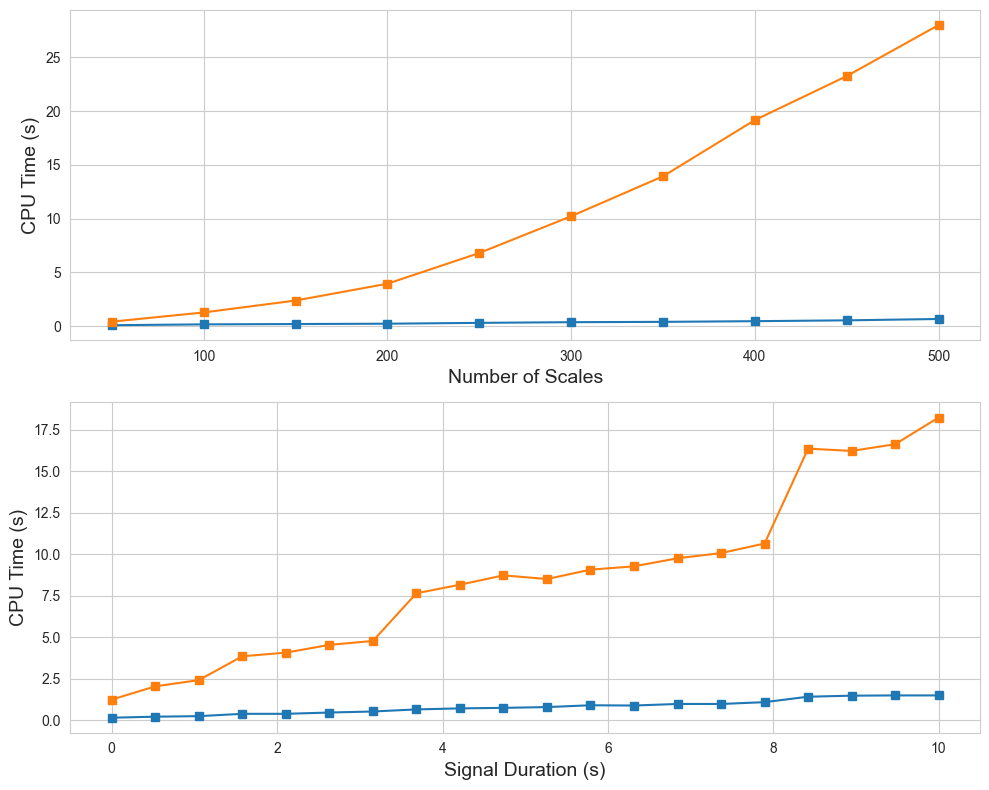}
    \caption{CPU-time comparison. Top: increasing frequency resolution; bottom: increasing number of samples. Legend: \textcolor{mplblue}{\rule{.5em}{.5em}} - Our model, \textcolor{mplorange}{\rule{.5em}{.5em}} - PyCWT benchmark.}
    \label{perf}
\end{figure}

\section{System Implementation}

To evaluate the Coherent Multiplex pipeline, a full simulation and analysis environment has been implemented. The following subsections describe each stage of the prototype, along with representative figures.

\subsection{Signal Generation}

Incoming data is procedurally generated to show signals that come in and out of covariance. 8 input signals are synthesized as linear combinations of sine waves with randomized amplitudes, frequencies, and phases. This approximates the behavior of non-stationary real-world signals. Coherence is artificially introduced between selected pairs by sharing frequency components or phase structure at randomized intervals.

\begin{figure}[!t]
\centering
\includegraphics[width=0.49\textwidth]{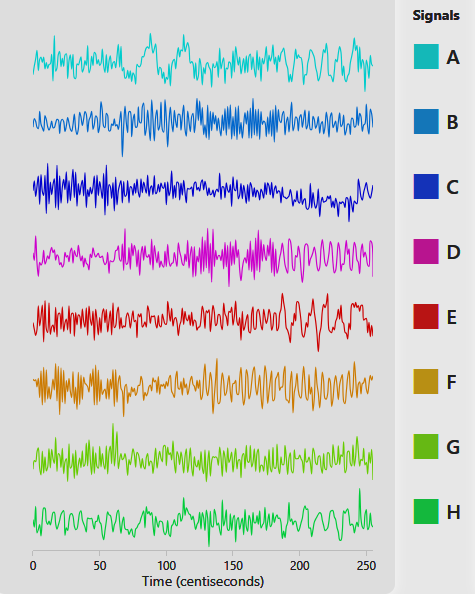}
\caption{8 signal streams are monitored over a rolling window. Signals are labeled A-H, and color coded for identification.}
\label{sigs}
\end{figure}

Signals are processed continuously using a rolling buffer, and maintain a window of 256 samples. The model is set to produce and advance by one sample each centisecond, and displays the signals as shown in Fig.~\ref{sigs}.

\subsection{Spectral Feature Extraction}

\begin{figure}[t]
\centering
\includegraphics[width=0.43\textwidth]{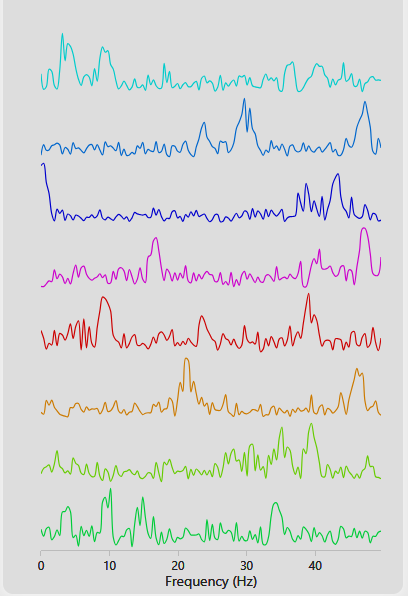}
\caption{Fourier magnitude spectra of each signal at a given time step. Pairwise cosine similarities are computed between rows.}
\label{freqs}
\end{figure}

For each update, the FFT is computed using a fast backend with pre-aligned reusable memory buffers. This implementation imports the FFTW C subroutine, which will enable scaling up. The magnitude spectra of the FFT results are extracted and used to compute pairwise cosine similarity. The real part of the FFT results are displayed in the frequency domain, as shown in Fig.~\ref{freqs}.

\subsection{Graph-based Similarity Modeling}

A fully connected network is made to represent the 8 signals as nodes, and the edges between nodes are weighted by the cosine similarity between the corresponding signals' frequency-domain vectors. This network serves as the first layer of analysis, allowing visualization of potential coherence relationships and identification of candidate pairs for further analysis.

\begin{figure}[htbp]
\centering
\includegraphics[width=0.4\textwidth]{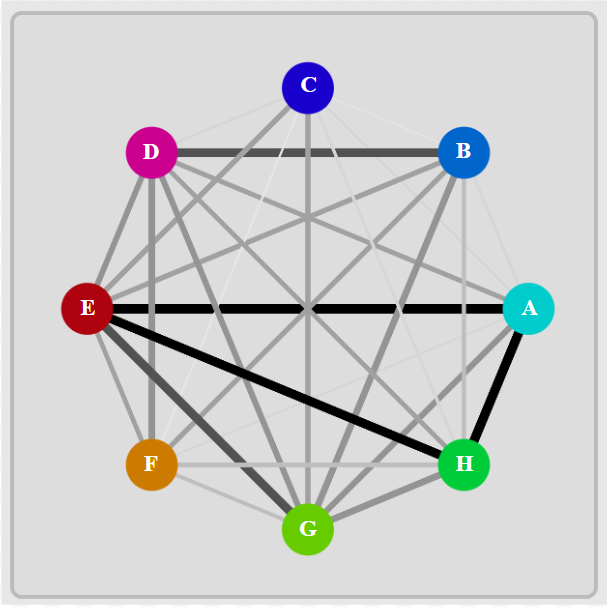}

\caption{Cosine similarity graph. Nodes represent signals; edge weights represent similarity. All possible pairs are connected. Node labels A through H begin at 3~o'clock and are indexed counter-clockwise.}
\label{graph}
\end{figure}

A wider, darker edge indicates a small angle between the two nodes that it connects. At the timestep that Fig.~\ref{sigs} was captured, the highest similarities were between pairs A-E, A-H, and E-H, which is reflected by the darkest edges linking nodes A, E and H in Fig.~\ref{graph}.

\subsection{Wavelet Coherence Analysis}

\begin{figure*}[htb]
\centering
\includegraphics[width=0.85\textwidth]{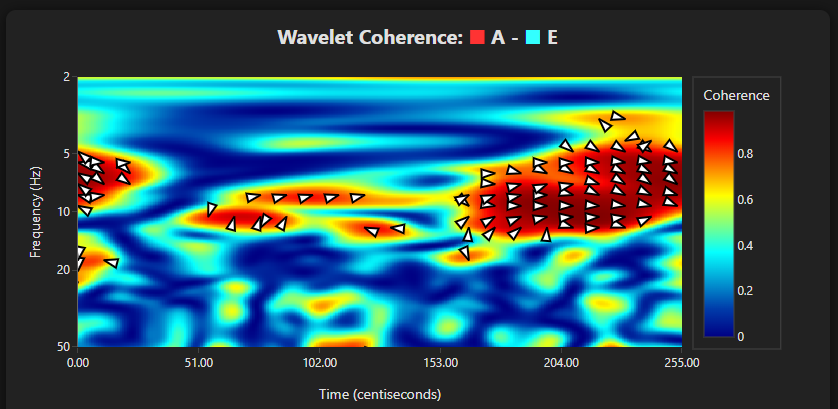}
\caption{Wavelet coherence analysis between selected signal pairs. High-coherence regions and phase arrow vectors are visible.}
\label{coh}
\end{figure*}

For deeper analysis in time and frequency relationships, wavelet coherence is selectively computed between the pair exhibiting the highest similarity at each update. Transform arrays are efficiently computed via the FCWT algorithm for each signal, and coherence is evaluated in the time-frequency domain. Visualizations in this implementation, like the one shown in Fig.~\ref{coh}, include phase arrows indicating lead/lag relationships, but ignore the cone of influence, which is overlaid on many coherence visualizations for masking of boundary effects.

\section{Discussion}
\subsection{Applications}
Wavelet coherence, while initially developed for studying long-term climate phenomena such as the El Ni\~no–Southern Oscillation and monsoon activity~\cite{torrence1998practical}, has been adopted across a wide range of disciplines. Notable applications include:

\begin{enumerate}

\item
\textbf{Economics:} Wavelet coherence is used to explore dynamic relationships among financial time series, including equities, foreign exchange rates, and macroeconomic indicators. It allows detection of transient co-movements and structural breaks that are not captured by stationary models.
\item
\textbf{Neuroscience:} The technique is widely applied to electroencephalogram (EEG) and magnetoencephalogram (MEG) data to assess time-varying functional connectivity between regions of the brain, aiding research into cognition, neurological disorders, and stimuli response.
\item
\textbf{Biomedical and Physiological Signal Analysis:} Coherence metrics are employed to evaluate interactions between biological signals such as heart rate variability and respiration, providing insights into autonomic regulation and disease progression.

\end{enumerate}

In these domains, the number of signals to be compared often exceeds the capacity of conventional coherence algorithms. For instance, EEG studies may involve 128 to 512 electrodes, while financial platforms such as New York Stock Exchange National maintain and update the prices of over 8000 listed securities. Traditional pairwise coherence analysis would be immediately ruled out on the basis of prohibitive computational expense.

The presented architecture overcomes these limitations by employing a computationally efficient multiplexing approach that scales to high-dimensional input. The system is adaptable to constrained hardware environments, and the current prototype operates effectively on a dual-core virtual machine with 3.5 GB RAM, demonstrating both portability and performance. These characteristics render the design suitable for real-world deployments in low-cost or embedded systems, as well as for high-throughput research applications.

\subsection{Future Work and Extensions}

The described system architecture is designed to support a wide range of modifications and extensions without compromising the integrity or performance of its core operations. Potential enhancements include the following:

\begin{enumerate}
  \item \textbf{Alternative Similarity Heuristics:} The cosine similarity metric can be substituted for alternative measures of signal correlation beyond cosine similarity, given computational allowance. Examples include Pearson correlation \cite{pearson1895}, phase-locking value (PLV) \cite{lachaux1999measuring}, MSC \cite{coherence_msc}, or Granger causality \cite{granger1969causality}.

  \item \textbf{Real-Time Alerting Mechanisms:} The framework can incorporate rule-based or model-based alert systems to trigger notifications when coherence metrics exceed or fall below predefined thresholds. These alerts can be used for anomaly detection, early warning systems, or to trigger downstream processing \cite{chandola2009anomaly}.

  \item \textbf{Substitution of Wavelet Coherence:} In use cases where scale-invariance is not required, the wavelet coherence module may be substituted with the cross-wavelet transform \cite{grinsted2004application}, short-time Fourier transform (STFT) \cite{cohen1995time}, or any comparable time-resolved transform method.

  \item \textbf{Stacked and Parallel Processing Layers:} The system can be extended with multiple sequential or parallel processing layers. For instance, wavelet coherence outputs may be jointly evaluated alongside macroeconomic indicators, trend models, or biomedical state variables to augment signal classification or correlation.

  \item \textbf{Adaptive Thresholding via Resource Monitoring:} Dynamic adjustments can be made to similarity thresholds based on monitoring of hardware resources such as CPU load, memory usage, or latency. This allows graceful degradation or scaling of analysis intensity to maintain responsiveness under constrained conditions.

  \item \textbf{Significance Testing:} An additional processing layer that implements statistical significance testing, such as through non-stationary surrogate data \cite{schreiber2000surrogate}, can be integrated. This layer would serve to robustly validate detected coherence patterns, improving the reliability of real-time inferences without affecting the core processing pipeline \cite{torrence1998practical, grinsted2004application}.

  \item \textbf{Neural Network Extensions:} Deep learning modules such as recurrent neural networks (RNNs) \cite{hochreiter1997long} or graph neural networks (GNNs) \cite{scarselli2008graph} may be appended to the multiplex to facilitate the automated interpretation, classification, and annotation of coherence dynamics.
\end{enumerate}

These modifications are intended to enhance system's adaptability across a range of application domains, including monitoring systems, embedded hardware deployments, and large-scale analytical platforms. The underlying modular design of the Coherent Multiplex framework ensures that such modifications can be integrated with minimal impact on system stability or core functionality.

\section{Conclusion}

We have presented the Coherent Multiplex, a scalable framework for analyzing wavelet coherence in large-scale multivariate time series. By combining spectral similarity filtering with wavelet coherence analysis within a graph-based architecture, the system efficiently captures dynamic relationships across signals. Its modular design ensures extensibility and adaptability, making it well-suited for diverse application domains such as neuroimaging, financial monitoring, and physiological signal tracking. Future work will focus on integrating significance testing, adaptive thresholding, and machine learning techniques to further enhance performance and reliability.

\bibliographystyle{IEEEtran}
\bibliography{IEEEsub}

\end{document}